\documentclass{emulateapj}

\begin{document} 
\title{\bf The Magellanic Clouds Photometric Survey: \break
The Large Magellanic Cloud Stellar Catalog and Extinction Map}

\author{Dennis Zaritsky, Jason Harris}
\affil{Steward Observatory, Univ. of Arizona, Tucson, AZ, 85721, email:
dzaritsky@as.arizona.edu} 
\author{Ian B. Thompson}
\affil{Carnegie Observatories, 813 Santa Barbara St., Pasadena, CA 91101,
email:ian@ociw.edu}
\author{Eva K. Grebel}
\affil{Astronomical Institute of the University of Basel, Venusstrasse 7, CH-4102 Binningen,
Switzerland, email:grebel@astro.unibas.ch} 

\begin{abstract}
We present our catalog of $U$, $B$, $V$, and $I$ stellar
photometry
of the central $64$ deg$^2$ area of the Large Magellanic Cloud.  Internal and 
external astrometric and photometric tests using
existing optical photometry ($U$, $B$, and $V$ from Massey's bright star
catalog and $I$ from the near-infrared
sky survey DENIS)
are used to confirm our observational uncertainty estimates.
We fit stellar atmosphere models to the optical data 
to check the consistency of the photometry for individual stars 
across the passbands and to estimate the line-of-sight
extinction. Finally, we use the estimated line-of-sight extinctions to produce
an extinction map across the Large Magellanic Cloud, confirm
the variation of extinction as a function of stellar population, and produce
a simple geometrical model for the extinction as a function of stellar population. 
\end{abstract} 


\section{Introduction}

A galaxy's star formation history is encoded within its stellar populations. 
Outside of our own galaxy, the Magellanic Clouds provide our most detailed view of 
galaxies that are still vigorously forming stars. 
We present the catalog of stellar photometric data
from the Magellanic Clouds Photometric Survey (MCPS)
for the entire Large Magellanic Cloud
(LMC) survey region (roughly $8.5^\circ \times 7.5^\circ$, with the longer direction
corresponding to the east-west axis).
Our data are either deeper, cover a wider
area, or include a larger number of filters (the 
inclusion of $U$ is particularly important
for studies of dust and young stellar populations) than available 
optical catalogs.

In addition to describing and providing the catalog, we construct and analyze extinction maps
of the LMC. As we demonstrated for 
a portion of the LMC \citep{z99} and for the SMC \citep{z02}, the extinction properties in
the Clouds vary both spatially and as a function of stellar population.
Therefore, for many scientific purposes the catalog alone is insufficient, one
must correct the observed magnitudes and colors for a complex
extinction pattern. 
We describe the MCPS in \S2, use the photometry
to generate extinction maps of the LMC for two different stellar
populations in \S3, model the relative distributions of young stars, old stars, and dust
in \S4, and summarize in \S5.

\section{The Data}

The data come from Magellanic Cloud Photometric Survey, originally described by 
\cite{zht97}. Using the Las Campanas Swope telescope (1m) and the
Great Circle Camera \citep{zsb96} with a 2K CCD, we obtained drift-scan images 
for both Magellanic Clouds in Johnson $U, B, V, $ and Gunn $i$. The effective
exposure time is between 3.8 and 5.2 min for LMC scans and the pixel scale
is 0.7 arcsec pixel$^{-1}$. Typical seeing is $\sim$1.5 arcsec and
scans with seeing worse than $\sim$ 2.5 arcsec are not
accepted. Magnitudes are placed on the Johnson-Kron-Cousins
photometric system \citep{l83,l92}.
Scan images from observing runs from October 1995 to December 1999 are
included in this catalog.  Additional pointed observations were performed in 
December 2001 to fill in for unacceptable data from earlier observations.
The data are reduced using a pipeline that
utilizes DAOPHOT II \citep{stet97} and IRAF\footnote{IRAF is
distributed by the National Optical Observatories, which are operated
by AURA Inc., under contract to the NSF}.  Only stars with both $B$
and $V$ detections are included in the final catalog.

Details of the reduction procedure, construction of the catalog, and 
quality assurance are presented by \cite{z02}. Because the SMC and LMC
data were taken concurrently, the quality of the two datasets is
identical. We present only a cursory description of the procedure here.
The pipeline for reducing individual scans is a fairly standard
application of DAOPHOT. Each of the 76 scans ($\times$ 4 for the four filters), 
which are 11000
pixels long and 2011 pixels wide, 
is divided into 11 by 2 subscans that are
roughly 1100 by 1100 pixels, with overlap between the subscans that enables
us to compare the results from the independent 
photometric reductions. 
The result of the reduction pipeline is a catalog 
of instrumental photometry for each detected star in each filter
and its right ascension and declination. We derive the astrometric solution 
from 
a comparison to stars in the Magellanic Catalogue of Stars (MACS; 
\citet{tu96}), whose coordinates are on the FK5 system.

We match the instrumental magnitudes of stars in different filters 
using a positional match that associates the nearest star on the sky
within an aperture that is 3 times either the positional rms of the 
astrometric standards in that
subscan or 1.2 arcsec, whichever is larger. The $V$ frame is used as
the reference and only stars that have a match in the $B$ frame are
retained for the final catalog.  In crowded areas it is possible that
the ``nearest'' star in one filter is not the correct match to the $V$
reference because of the uncertainties in the astrometric solution. 
We see some evidence of this problem when comparing to
other data and when fitting atmospheric models (we find stars with highly
anomalous colors). Except near the faint limit of the catalog or in extremely
crowded regions, this issue appears to be a minor problem.  
These errors can be estimated reliably using artificial star
simulations.

\begin{figure}
\plottwo{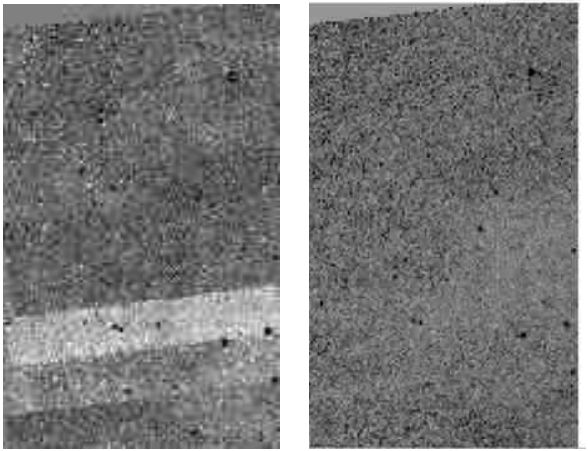}{Zaritsky.Fig1b.epsf}
\figcaption{
A section of the red clump stellar density map for the LMC. The left panel contains
the I-band photometry with no corrections, the right panel contains the photometry
after correction. 
\label{rcfix}}
\end{figure}

One significant difference bewteen the photometric solution of the LMC and SMC 
data, is that for the LMC data we do not automate the adjustment of the
scan-to-scan photometry. We find that the photometry among scans is generally 
consistent, and only a few manual adjustment of scan zeropoints were
necessary. These were implemented to remove obvious discrepancies 
observed in a map of the density of red clump stars (for an example see Figure \ref{rcfix}).
The adjustment of specific individual scans avoids the potential for systematic
drift present in an automated procedure.

We present the catalog of astrometry and photometry for 24,107,004 stars 
as an ASCII table (see Table 1 for a sample).
Columns 1 and 2 contain the right ascension and declination (J2000.0)
for each star. Columns 3-10 contain the pairings of magnitudes and
uncertainties for $U$, $B$, $V$, and $I$ magnitudes.  The last column is 
a quality flag that is described in Table 2, based on comparisons to 
Massey's (2001) catalog and our fitting of the spectral energy distribution (\S3).
A $V$ band stellar density map of the LMC
constructed from the 
catalog using stars with $16 < V < 20$ is shown in Figure~\ref{lmcimagev}.
The digital catalogs allow one to make analogous images for a variety of populations
(for an example of the distribution of young and old stars in the LMC, as done for the
SMC by \cite{z00},
see Figure 2).

\begin{deluxetable*}{rrrrrrrrrrr}
\tablecaption{The MCPS Catalog for the Large Magellanic Cloud \tablenotemark{A}}
\tablehead{
\colhead{RA}&\colhead{Dec}&\colhead{$U$}&\colhead{$\sigma_U$}&\colhead{$B$}&\colhead{$\sigma_B$} &\colhead{$V$} & \colhead{$\sigma_V$}&\colhead{$I$}&\colhead{$\sigma_I$}&\colhead{Flag}\\
}
\startdata 
4.490203&$-$72.22987&.000&.000&20.740&.060&19.746&.054&18.535&.063&0\\
4.490215&$-$72.19868&.000&.000&22.694&.271&22.788&.310&.000&.000&0\\
4.490216&$-$72.28524&.000&.000&21.646&.099&20.894&.071&.000&.000&0\\
4.490218&$-$72.37505&.000&.000&21.455&.107&21.050&.086&20.795&.167&0\\
4.490222&$-$72.29113&.000&.000&21.379&.076&21.298&.106&.000&.000&0\\
4.490228&$-$72.09232&.000&.000&21.668&.104&20.700&.067&20.114&.087&0\\
4.490231&$-$72.09853&21.724&.410&21.221&.078&20.740&.062&20.015&.099&0\\
4.490231&$-$72.17493&.000&.000&22.277&.202&22.220&.213&21.657&.271&0\\
4.490236&$-$72.18102&20.326&.168&20.013&.070&19.134&.078&18.055&.073&10\\
4.490241&$-$72.09316&21.588&.384&22.491&.231&22.009&.176&.000&.000&0\\
4.490252&$-$72.13191&.000&.000&23.283&.378&22.997&.356&.000&.000&0\\
4.490256&$-$72.13083&20.287&.140&20.379&.061&19.989&.055&.000&.000&0\\
4.490261&$-$72.20712&.000&.000&22.687&.318&23.412&.495&.000&.000&0\\
4.490273&$-$72.18983&.000&.000&22.368&.201&21.895&.135&21.320&.271&0\\
4.490280&$-$72.40957&.000&.000&21.811&.120&21.296&.113&.000&.000&0\\
\enddata
\tablenotetext{A}{The complete version of this table is in the electronic
edition of the Journal. The printed edition contains only a sample.
}
\label{catalog}
\end{deluxetable*}

\begin{figure}
\plottwo{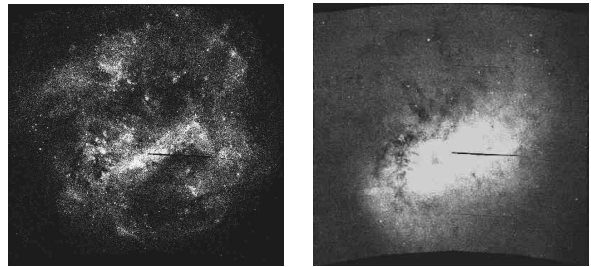}{Zaritsky.Fig2b.epsf}
\figcaption{
The stellar density map of the LMC for 
the young population ($V < 18.5$ and $-0.2 < B-V < 0.3$) in the left panel and
an old population ($V < 19.1$ and $0.7 < B-V < 1.5$) in the right panel. 
North is up and East is to the left. The image shows the entire $\sim 8.5^\circ
\times 7.5^\circ$ survey region.
\label{lmcpops}}
\end{figure}

\begin{deluxetable}{lr}
\tablecaption{MCPS Catalog Flags\tablenotemark{A}}
\tablehead{
\colhead{Description}&\colhead{Value}\\
}
\startdata 
Replaced with Massey's (2001) photometry \& astrometry&1\\
Colors successfully fit with stellar atmosphere model&10\\
Colors poorly fit with stellar atmosphere model&20\\
\enddata
\tablenotetext{A}{The flag values are additive. For example, a 
star with a flag value of 11 has had its photometry replaced with
that of Massey's (2001) catalog and is fit well with the
atmospheric models.
\label{flags}
}
\end{deluxetable}

The magnitude limit of the survey varies as a function of stellar
crowding. We find little visible evidence for incompleteness at $V <
20$ (Figure~\ref{lmcimagev}), but the scan edges and different scan sensitivities
become visible when
plotting the stellar surface density for stars with $20 < V < 21$
(Figure~\ref{incompletev}). 
The $U$ and $I$ data are incomplete at brighter magnitudes than the
$B$ and $V$ 
data. The $U$ and $I$
photometry, even in sparse areas, is severely incomplete below $U=21.5$ and
$I = 22$ (comparable limits in the two other
bands are $B=23.5$ and $V= 23$). Any statistical analysis of this catalog
fainter than $V < 20$ requires artificial star tests to determine incompleteness, 
which is becoming significant at these magnitudes.

\subsection{Photometric and Astrometric Accuracy}

We presented extensive tests of the photometry and astrometry of the
MCPS in \cite{z02}. However, we revisit some of those here
for completeness. One particularly useful internal test is based
on a map of the red clump mean luminosity. We calculate the
mean magnitude of the red clump in 50 $\times$ 50 arcsec boxes over
the survey region. Although large-scale variations in the mean magnitude
may truly exists (for example due to a tilt of the LMC relative to a
constant-distance surface; see 
\citet{vdm01} for a demonstration of an analogous effects with the red giant branch
tip magnitude),
any localized variation, in particular one that traces
scan or subscan boundaries, reveals a problem region.
In Figure~\ref{clumpmos} we show the maps of the deviations from the mean
red clump magnitude, for each
filter. There are various important features in these panels. First,
all four panels show increased noise toward the edges because there
are fewer clump stars at large radii from the LMC and contamination
for foreground Galactic stars is proportionally greater. Second, the
$U$ and $I$ band frames show more structure that follows scan 
and subscan edges. Although we interactively corrected the most egregious of
these (\S2), we did not correct subscans within the bar region nor did we apply
correction in cases where the correction needed was ambiguous. Recall that
the $U$ and $I$ images are not as deep as the $B$ and $V$ images,
so that some of the differences could be due to senstivity and completeness
variations among scans (the clump stars are near the $U$ magnitude limit).
Third, there are large spatial scale variations that trace the bar, and repeat
in the various colors. These are real variations in the mean magnitude 
of clump stars, presumably either due to physical effects (geometry, population
differences, or variable extinction).
We conclude that with the exception of some regions in the $U$ and $I$ band
data, the photometry is self consistent among scans. The worst variations seen in 
the $U$ band correspond to $\sim$ 0.1 mag.

\begin{figure}
\plotone{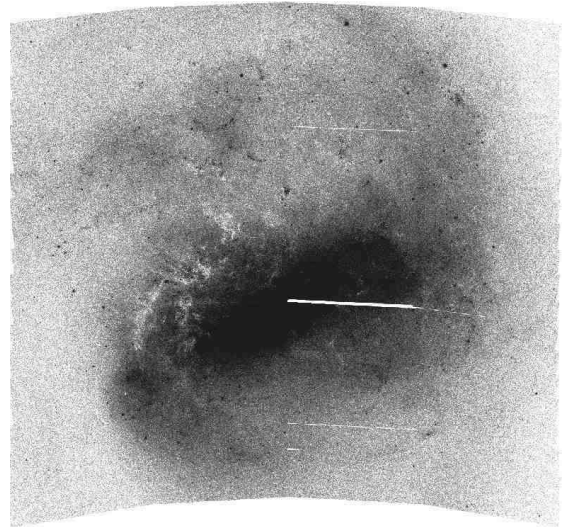}
\figcaption{
The stellar density map of the LMC for stars with $16 < V < 20$. 
North is up and East is to the left. The image shows the entire $\sim 8.5^\circ
\times 7.5^\circ$ survey region. The 
nearly-horizontal white lines are caused by slight gaps between scans, which
occurred if either the $V$ or $B$ scan was slightly misaligned. The irregular patchiness
is due to extinction that is confirmed in \S3.
\label{lmcimagev}}
\end{figure}

In \cite{z02} we compared our photometry to various existing catalogs. 
The quality of the photometry presented here is identical since the LMC
and SMC data were taken concurrently. However there is one issue that
requires further comment. 
\citet{massey02} has produced a photometric catalog ($U, B, V$
and $R$) of bright stars in the LMC.
In particular, because of his interest in the 
upper main sequence, he pays acute attention to obtaining
accurate $U$-band photometry for young LMC stars, 
which is often difficult because of the
lack of blue ($U-B < 0.0$) standards.  He found that the combination of
filter, CCD passbands, and the gravity sensitivity of the Balmer jump
produced rather complex color transformation between his $U$-band
observations and the standard Johnson system. Specifically, he required
a second order color-term to correct his magnitudes.

We had previously used his uncorrected photometry to correct our 
$U$-band photometry (a slightly linear color term was added to place
our photometry on his system, which was more extensively calibrated
than our $U$-band system; \cite{z02}). However, between that correction and 
his publication of the catalog, he identified the additional complication
in the calibration of the $U$-band data mentioned above. Because the effect appears to 
arise from the
application of a calibration based on dwarf stars to the supergiants 
at these magnitudes in the LMC, our expectation is that
our published photometry is correct for dwarfs, but that 
the second order color term
is required for supergiants. Because most of the stars in the catalog are not
blue supergiants, our general results should be unaffected.
This conclusion is supported indirectly by
the results of our modeling of the SMC population, where we found
good agreement with the overall photometry and models
\citep{hz03}. Nevertheless, detailed photometry of the upper main sequence
is more uncertain than quoted in our original catalog \citep{z02}.

To examine this issue further, we match our photometry of LMC stars to that of 
\cite{massey02}.
We match stars brighter than $V = 15$ in Massey's catalog to stars in
the MCPS by finding the star in the MCPS within 7.5 arcsec 
that has the closest $V$ magnitude, as was done for the SMC data. The selection
of the search radius is a compromise between assuring that the search finds the
corresponding star while moderating the incidence of false matches. Because
the bright star catalog has a relatively low density on the sky, we use 
a large search radius.
As expected, we find that our uncorrected $U$ magnitudes have a strong color
term relative to those of Massey (see Figure \ref{masseycolor}). 
We can correct our magnitudes by roughly fitting
a second order $B-V$ color term, which produces the second panel in Figure \ref{masseycolor}.
The correction to our $U$ photometry  is
$\Delta U = -0.5(B-V -0.7)^2  + 0.23$ if the $(B-V) < 1$ and $\Delta U = 0.05$ 
otherwise. The panels on the right hand side of the 
Figure demonstrate that the correction does indeed decrease the scatter between
the two sets of measurements, but the systematic error is  still as large
as 0.1 mag at certain values of $B-V$.
Given the uncertainty of which stars the correction should be applied to due to the
gravity sensitivity, we
have chosen to not apply the correction to our data as presented in the catalog. 
However, we will present results in the discussion of derived extinctions with and 
without the correction. We find that whether the correction is applied or not does not affect 
our results.

\begin{figure}
\plotone{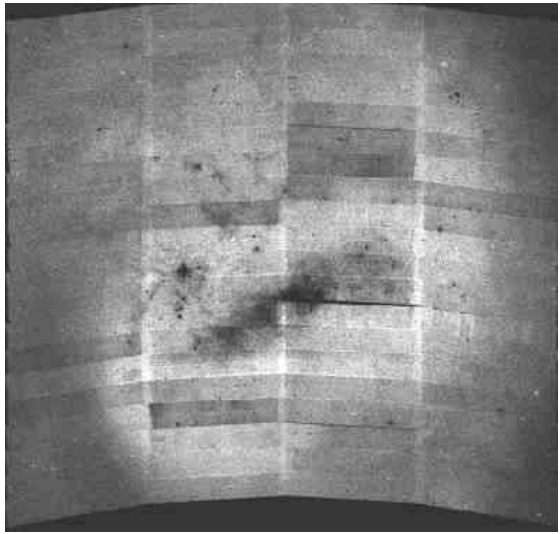}
\figcaption{Stellar density map of stars with $20 < V < 21$. Various signs
of incompleteness and double detections that were unmatched 
are evident. The scan-to-scan variations
reflect varying completeness limits due to seeing variations. Bright regions
of overlap between frames show the incidence of false detections. The dark
patches along the bar and the region of 30 Doradus show crowding compleness limits.
\label{incompletev}}
\end{figure}

Stars in
our catalog that are brighter than 13.5 in $B$ or $V$ are prone to
substantial photometric uncertainty \citep{z02}.  In the SMC catalog, 
we replaced the
photometry and astrometry for stars brighter than this
limit with those of \citet{massey02}. However, the Massey catalog for the LMC
covers only about a quarter of the area that ours covers. Therefore, most
of our stars brighter than 13.5 in $B$ or $V$ have suspect photometry and
astrometry. We have, when possible, replaced our 
photometry
with Massey's, applying the mean photometric offsets found between the
MCPS and Massey's catalog, to Massey's
data. 
To indicate which
stars do have corrected photometry the catalog includes a flag (see
Table 2 for description of the quality flags). The photometry and
astrometry of 10912 stars are corrected using Massey's catalog. 

\begin{figure}
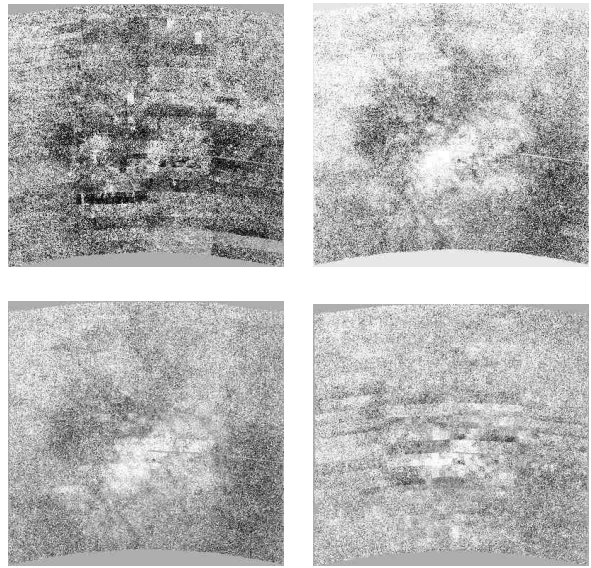

\plottwo{Zaritsky.Fig5a.epsf}{Zaritsky.Fig5b.epsf}
\vskip 12pt
\plottwo{Zaritsky.Fig5c.epsf}{Zaritsky.Fig5d.epsf}
\figcaption{Comparison of mean red clump magnitudes in U (top left), B (top right), V
(bottom left), and I (bottom right).
Greyscale ranges from $-$0.3 to 0.3 mag. The entire survey region is shown.
\label{clumpmos}}
\end{figure}

\begin{figure}
\plotone{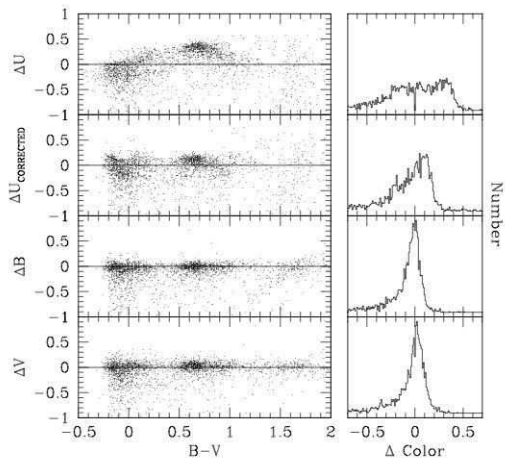}
\figcaption{The comparison of magnitudes 
as a function of stellar color 
between the Massey survey and the MCPS. The upper panel shows the results
from the original $U$ band photometry, the second panel shows the
corrected $U$ band photometry. 
\label{masseycolor}}
\end{figure}

\begin{figure}
\plotone{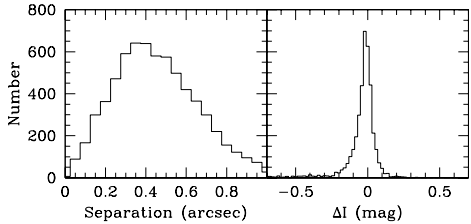}
\figcaption{The comparison of DENIS $I$ band catalog
with our data. The panel in the left illustrates the astrometric
differences for matched stars with $I$-band magnitudes. The panel in the right
shows the distribution of differences in $I$ magnitudes between the two surveys.
\label{deniscomp}}
\end{figure}

Finally, we compare our $I$-band photometry to that in the DENIS
catalog. Although the DENIS catalog is primarily an IR catalog, it
contains an $I$ band channel and \citet{c00} present a
point-source catalog in the regions of the Magellanic Clouds.
We use a search aperture of 3.5 arcsec for matches (smaller than
that for comparison to the bright star catalog because of the increased
stellar density).
The distribution of astrometric and photometric differences for
matched stars are plotted in Figure~\ref{deniscomp}. In agreement with
our previous results, we find that the astrometric
accuracy is subpixel for the majority of the matches. The mean
difference is 0.45 arcsec.  For
matched stars whose magnitude agree to within 1 magnitude, the mean photometric
difference between the two surveys is $0.01$ mag.

\section{Extinction Properties}

A comparison between stellar atmosphere
models and observed colors can be used to infer the extinction toward
individual stars \citep{gr,z99}.
We use the models of  \cite{l97} and our $U, B, V$, and $I$ photometry to measure the
effective temperature, $T_E$, of the star and the line-of-sight extinction,
A$_V$, in the manner described by \cite{z99}.
We adopt a standard Galactic extinction curve \citep{schild}, which
is acceptable for the LMC over the optical wavelength region (see \cite{gordon}).
The model fitting is least
degenerate between $T_E$ and A$_V$ for stars with derived
temperatures in the ranges $5500 {\ \rm K}\le T_E \le 6500 {\ \ \rm K}$ and
$12,000 {\ \rm K}\le T_E \le 45000 {\ \rm K}$.  Therefore, 
we construct A$_V$ maps of the LMC
from the line-of-sight A$_V$ measurements to the set of ``cool''
stars ($5500 {\ \rm K}\le T_E \le 6500 {\ \rm K}$) and the set of ``hot''
stars ($12000 {\ \rm K}\le T_E \le 45000 {\ \rm K}$) with good quality
photometry in all four filters ($\sigma_U < 0.4$, $\sigma_B < 0.2$, $\sigma_V < 0.2$,
$\sigma_I < 0.2$) and good model fits ($\chi_\nu^2 < 3$).
In addition
to these criteria, we imposed a reddening-independent magnitude cut
($V < 19.0 + 3.2 \times (B-V)$). We caution however, that although the
magnitude cut is reddening-independent, there is a bias in the catalog
against highly extincted stars simply because the catalog itself is
magnitude limited.
Stars which satisfy the photometric criteria
and have an acceptable model fit ($\chi_\nu^2 < 3$) have a quality flag of
$10$ and those that do not have an acceptable model have a quality flag of $20$
in our catalog (Table \ref{flags}).

\begin{figure*}
\plottwo{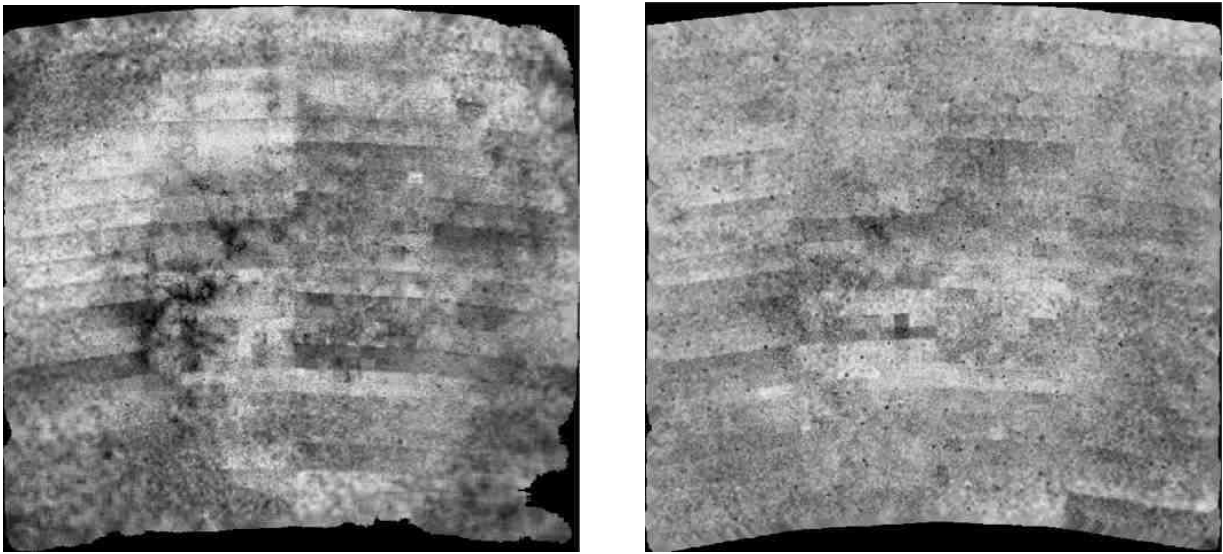}{Zaritsky.Fig8b.epsf}
\figcaption[extinctionmap.ps]{The spatial map of extinction
values as derived for both the cooler, older stars ($5500 {\ \rm K}\le T_E 
\le 6500 {\ \rm K}$ ; 
right panel) and the hotter, younger stars ($12000 {\ \rm K}\le  T_E
\le 45000 {\ \rm K}$; left panel) for the entire survey region. The small localized circular regions of apparent high extinction are
globular clusters, which have some anomalous photometry because of their
high stellar densities. Sharp spatial variations are due to scan-to-scan 
photometry differences of a few hundredths of a magnitude. 
\label{extinctionmap}}
\end{figure*}

In Figure ~\ref{extinctionmap} 
we show the spatial distribution of line-of-sight extinction derived from both the hot and cold
stellar populations\footnote{The extinction catalog is available for query through the MCPS
home page at http://ngala.as.arizona.edu/dennis/mcurvey.html}.
Because the recovery of A$_V$
is quite sensitive to color, subtle differences in the scan
photometry are highlighted in the extinction maps.
For example, a set of small photometric
differences (0.03 mag in opposite senses in $B$ and $I$, 
so that $B-I$ has changed by 0.06 mag) creates an extinction discontinuity
of the magnitude observed at many scan edges in the hot population map of Figure ~\ref{extinctionmap}.

The principal coherent extinction structure within the LMC is the 
increase in extinction in the hot star population along the north-east ridge of
the LMC bar. This structure is also visible in the map from the colder
stars. As expected, the increased extinction correlates with sites of star formation
that can be seen in the stellar density images (Figure ~\ref{lmcpops}), 
in 100$\mu$m images \citep{wheelock}, and in CO maps \citep{fukui}. 

The A$_V$ histograms of the two populations are shown in
Figure~\ref{dust}. As we found for a small region of the LMC \citep{z99},
the mean extinction is lower for the cooler
populations (average A$_V$'s of 0.43 mag vs 0.55 mag for the cold vs. hot population,
respectively).  The bimodal distribution of extinction values among the cooler stars
is characteristic
of a geometry where the stars are distributed in front and behind a thinner
mid-plane dust plane (see Figure 12 of \cite{hzt}). The lack of such bimodaility
in the $A_V$ distribution of the hotter stars suggests that those are distributed within
the dust layer.  This geometric model is further supported by the correspondence between
the $A_V$ values of the maximum in the hot star $A_V$ distribution and the local minimum
in the cold star $A_V$ distribution.

\subsection{A Simple Model}

To explore the dependence of the $A_V$ distribution on the relative distributions
of the dust and stars, we develop a simple geometrical model. We construct a galaxy
using three components: a ``cold" star component that matches the projected
spatial distribution of such stars, a ``hot" star component  that matches the projected spatial 
distribution of such stars, 
and a dust layer that corresponds to the dust map
developed from the ``hot" stars plus additional components for a diffuse, homogeneous internal
dust layer and Galactic foreground, which is estimated to be minimal ($A_V \sim 0.05$).
The two dust components associated with the LMC, a clumped component seen in 
the extinction map and a possible smooth component that is not as easily detected, 
are defined to have the same vertical scale height.
We then randomly place ``cold" stars along the line of sight, but spatially constrain them
to follow the observed distribution by placing them at the same $(\alpha , \delta)$ as
each observed ``cold" star. Each star is reddended according to the local
value of $A_V$ (taken to reflect the midplane value of extinction as derived from the
hot stars) and its position
relative to the dust layer (for example, stars in front of the disk layer are extincted only by
the Galactic foreground component, while stars behind the dust layer are extincted
by twice the measured value of $A_V$ plus the hypothesized diffuse layer.

\begin{figure}
\plotone{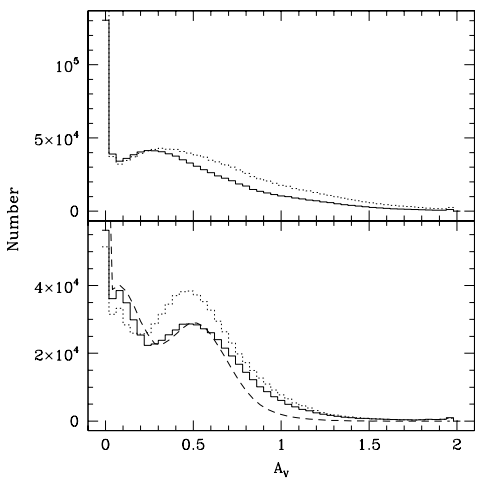}
\figcaption[dust.ps]{The distribution of extinction
values for the two stellar populations and a simple model.
We show the distribution of
$A_V$ for the 
hotter stellar population (upper panel) and 
for the colder stellar population (lower panel),
where the solid and dotted
line represent the unadjusted and adjusted $U$-band photometry.
The dashed line in the lower panel represents the expected distribution of $A_V$ for 
an older stellar populations derived from the simple model detailed in the text. The sharp
rise at zero extinction is due to observational uncertainties that scatter measurements
to negative $A_V$, which we set to $A_V=0$. The model shows the same rise
at $A_V = 0$.
\label{dust}}
\end{figure}

We explore a range of relative thicknesses of the star and dust layers, and of
the optical depth of the diffuse layer. A model that reproduces the salient features of
the distribution of $A_V$ for the cold population is shown in Figure \ref{dust}. The particular
model shown has a scaleheight for the cold stars that is ten times that of the dust,
a Gaussian distributed error in the measured $A_V$ of 0.125 mag, an optical depth
for the diffuse layer,  $\tau_f$, of 0.28 (if the optical depth as estimated by the young stars in a region
is $<$ 0.28, we set the value to 0.28), and a foreground extinction optical depth of 0.05,
which sets the position of the peak of the low-$A_V$ population. 
The model excellently reproduces
the $A_V$ distribution for $A_V < 0.7$ and underestimates $A_V$ for larger values.
It may therefore be the case that there are pockets of higher optical depth in the midplane
than modeled or a radial dependence in the extinction of the diffuse component that
perhaps increases toward the LMC center,  
but adding those in an ad hoc manner to the models
introduces far too many free parameters. Because of this systematic issue, we do 
not provide ``best-fit" values for the parameters of our model, but instead simply show
one satisfactory example. 

\begin{figure}
\plotone{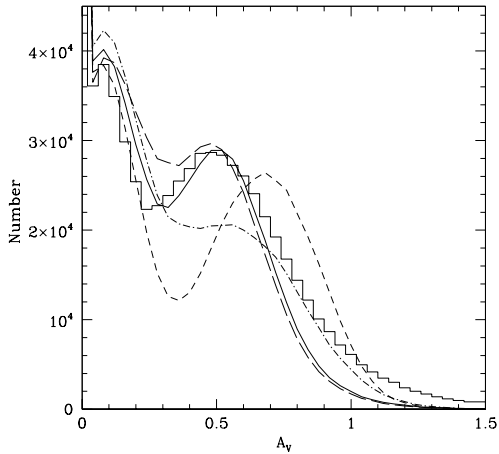}
\figcaption[dust1.ps]{A comparison of the distribution of extinction
values for the older stellar population and models.
The histogram shows the data, the solid curve shows the model shown in Figure \ref{dust},
the short-dashed line represents the $\tau_f = 0.4$ model, the long-dashed line
represents the model with a stellar scaleheight that is 5 times that of the dust layer,
the dot-dahsed line represents the exponentially declining $\tau_f$ model.
\label{failures}}
\end{figure}

There is some degeneracy between the various parameters,
but large deviations from the values quoted above produce qualitatively inferior fits. 
For example, modifying $\tau_f$ to correspond to $A_V=0.4$
produces a highly inferior fit (Figure \ref{failures}) and decreasing 
the stellar scaleheight by a factor of two fills in the valley between the 
two peaks of extinction values. An interesting variant of the model is to 
presume that $\tau_f$ has a radial dependence. If $\tau_f$ decreases exponentially
with increasing radius, we find that for small values of the scalelength 
(for example, a 1 kpc scalelength for the model
shown in Figure \ref{failures}), the second peak is diluted, due to the spread in
midplane extinction values in such a model, but that we do better at reproducing the
high end tail of the extinction distribution because we can drive the extinction
at the center of disk high and not affect large numbers of stars. Of course there
are other solutions to the high-end extinction distribution, so this agreement alone
does not argue for the model with exponentially declining $\tau_f$. Models
with large scalelengths approach the uniform $\tau_f$ model, while models with
smaller scalelengths further dilute the second peak.

While the distribution of dust is undoubtedly more complicated than the description
adopted in these models, the models are able to reproduce the peak at $A_V = 0$,
the peak at $A_V \sim 0.05$, the bimodal peak distribution, the position of the second
peak, and the tail toward high $A_V$ values with fairly minimal model assumptions.
This suggests that as a global average the model is qualitatively correct. Even so,
it is a much more complicated description of the internal absorption in galaxies
than generally adopted. Some studies that focus on internal extinction \citep{witt, misselt}
explicitly deal with the distributions of stars and dust, but the general default correction is
based on a foreground sheet assumption and an effective extinction curve. Because of
the partial correlation of dust with star formation, this effective extinction curve is
likely to be highly complex, dependent on geometry, and sensitive to the evolutionary
state of the system.

\section{Summary}

We have conducted a broad-band photometric survey of the Magellanic
Clouds.  We present the data for over 24 million stars in
the $8.5^\circ \times 7.5^\circ$ survey area centered on the Large
Magellanic Cloud.  The catalog contains positions
(right ascension and declination in J2000 coordinates) and $U$, $B$, $V$, and $I$ magnitudes
and uncertainties in the Johnson-Kron-Cousins photometric system measured 
from our drift scan images. 

Using this catalog, we have constructed extinction maps for two
stellar populations in the LMC.
We find 1) that this dust is highly localized near the younger, hotter stars,
and in particular toward regions immediately east and northeast of the center
of the LMC,
2) that aside from these regions of higher extinction, there is no discernible global
pattern, 3) that on average the extinction
toward  the younger, hotter stars is only about 0.1 mag larger, but that the distributions
of extinctions are entirely different, 4) that the distribution of extinctions along lines-of-sight
toward the older stars is bimodal, and 5) that the bimodial $A_V$ distribution is easily modeled
as stars in front and behind a thinner dust layer.
The two external galaxies
for which we now have highly detailed maps of extinction as a function
of stellar population both show significant differences in the 
extinction toward those populations. This difference, or at least the
potential for this difference, should be considered when correcting
the photometry of other galaxies for internal extinction. 

\vskip 1in
\noindent
ACKNOWLEDGMENTS: DZ acknowledges financial support from an
NSF grants (AST-9619576 and AST-0307482), a NASA 
LTSA grant (NAG-5-3501), and fellowships from
the David and Lucile Packard Foundation and the Alfred P. Sloan
Foundation. EKG acknowledges support from NASA through 
grant HF-01108.01-98A from the Space Telescope Science Institute and
from the Swiss National Science Foundation through grant 200021-101924/1.

\end{document}